\documentstyle[sprocl,cite,amssymb]{article}
\newcommand{\toline}[1]{}

\newcommand{\xtra}[1]{{.}}
\renewcommand{\xtra}[1]{{, \tt hep-th/#1}}

\newcommand{\xtrac}[1]{{.}}
\renewcommand{\xtrac}[1]{{, \tt cond-mat/#1.}}

\newcommand{\mathematica}[1]{{}}

\newcommand{\mm}[2]{{\vphantom{\vbox to 6mm{}}}}

\newcommand{\beqcol}{\begin{array}{rcl}}
\newcommand{\eeqcol}{\end{array}}

\newcommand\eq{\begin{equation}}
\newcommand\en{\end{equation}}

\newcommand{\One}{{\hbox{{\rm 1{\hbox to 1.5pt{\hss\rm1}}}}}}
\renewcommand{\One}{{\mathbb 1}}
\renewcommand{\One}{{\rm 1\!\!1}}

\newcommand{\opnup}[1]{\renewcommand{\\}{\\[50 pt]}}

\begin{document}

\title{\uppercase{Boundary integrable quantum field theories}}

\author{\uppercase{Patrick Dorey}}

\address{Department of Mathematical Sciences, University of Durham,\\
 Durham DH1 3LE, UK\\
E-mail: {\tt p.e.dorey@dur.ac.uk}}

%%%%%%%%%%%%%%%%%%%%%%%%%%%%%%%%%%%%%%%%%%%%%%%%%%%%%%%%%%%%%%

\maketitle

\abstracts{
A brief survey of recent results in the study of boundary integrable quantum field
theories, indicating some currently open problems.  Based on lectures given at the
2000 E\"otv\"os Summer School in Physics on ``Nonperturbative QFT methods and their
applications".
{~}\hfill DCTP/01/13; {\tt hep-th/0101174}}

%\section*{Introduction}
%
These lectures
concerned the properties of quantum field theories in the presence of boundaries. 
There are many different approaches to this subject. One can begin by studying 
{\it conformal}
field theories with boundaries -- the principal theme of the lectures at this
school by Jean-Bernard Zuber
and by Christoph Schweigert -- and then, as described in G\'erard Watts' lectures, 
consider their 
perturbations. In many cases these perturbations result in massive integrable quantum
field theories, and it was the direct study of such theories in their own right that 
formed my main topic.
A number of reviews of this subject can be found on the electronic
archives, and so in this
contribution I shall restrict myself to an outline of the questions touched on in my
talks, and a brief list of references to which the interested reader can turn
to find at least some of the answers.

The focus will be on boundary field theories which are {\em integrable}, and
if the usual locality conditions are also imposed then, just as for theories
without boundaries, the dimensionality of space must be restricted
to one. With the time dimension remaining infinite, there are 
then just two possible `boundary geometries': the theory can be defined either

\smallskip

$\bullet$ on a half-line
\begin{picture}(200,10)(0,0)
\thinlines
\put(33,-10){$\scriptstyle -\infty$}
\put(158,-10){$\scriptstyle 0$}
\put(40,0){\line(1,0){120}}
\put(160,-3){\line(0,1){6}}
\end{picture}

\smallskip

\noindent or

\smallskip

 $\bullet$ on a finite interval
\begin{picture}(200,10)(15,0)
\thinlines
\put(48,-10){$\scriptstyle a$}
\put(140,-10){$\scriptstyle b$}
\put(50,0){\line(1,0){92}}
\put(142,-3){\line(0,1){6}}
\put(50,-3){\line(0,1){6}}
\end{picture}

\bigskip
\medskip

It can then be studied either as a \underline{classical}, or directly as a
\underline{quantum} field theory. 
Key questions that one might ask include the
following:
\smallskip

\noindent{\bf (a)} 
For a given $\left\{\small{\mbox{classical}\atop\mbox{quantum}}\right\}$ 
integrable model on the full line (`in the bulk'), which boundary conditions are 
compatible with integrability?

\noindent{\bf (b)} 
Given a massive boundary integrable quantum field theory, how do bulk particles 
scatter off the boundary, and how should the `exact S matrix' technology be
generalised to encompass boundary problems?

\noindent{\bf (c)}
Can perturbation theory be set up to test any exact predictions?

\noindent{\bf (d)}
In the presence of a boundary, the spectrum of bulk excitations may, depending on
the boundary condition, be augmented by a number of `boundary bound states'. How
is this spectrum encoded in the amplitudes for the scattering of bulk particles off the
boundary?

\noindent{\bf (e)}
What does the spectrum of the theory look like on a finite interval?

\smallskip

Before any of these issues can be addressed, the properties of massive integrable
quantum field theories on the full line should be understood. The classic
reference is the article \cite{zamzam} by Zamolodchikov and
Zamolodchikov; a couple of more recent reviews, containing many more references, are
\cite{Ma,Da}.
\smallskip

Having digested this material, we can return to the novel questions which arise
when the model has a boundary. Much of the recent interest in this topic can be
traced to the pioneering
work \cite{GZa} by Ghoshal and Zamolodchikov. In particular, 
question {(b)}, concerning the correct generalisation of the ideas of
exact S matrix theory to boundary theories, was answered in reasonably complete
detail in this paper. Other early explorations of this point can be found in
references \cite{Ca,FKa,Sa}.

Questions {(a)} and (d) were also raised in \cite{GZa}. For the case of the
sine-Gordon model, it was argued on the basis of low-lying conserved quantities
that there should be a two-parameter family of classically integrable boundary
conditions for each bulk theory. 
The generalisation of this work to other classical theories, and the
development of more sophisticated techniques allowing complete integrability to
be established, provides a rich source of open problems. Further work
in this area includes \cite{CDRSa,Mca,CDRa,BCDRa,Ia}.
To treat question (d),
the spectrum of boundary bound states, one has to learn how to interpret
the pole structure of the reflection amplitudes. The basic rules were laid out in
\cite{GZa}, while in \cite{DTWa} it was pointed out that the story is sometimes
complicated by multiple rescattering processes, the so-called boundary
Coleman-Thun mechanism. These methods have been applied to a variety of models
\cite{SSa,DGa,MDa}, but it remains an open problem to place them on a solid
field-theoretic base.

This leads naturally to
question (c), the perturbative treatment of boundary models, and this
remains a difficult problem. A
framework for calculations has been set up in \cite{Ka,Cb,Ta}, and tested in a
number of examples. Particularly tricky are cases involving expansions about
non-trivial classical backgrounds; see
 \cite{CDRa,SSWa,Ba,PBa,Dea} for some sample discussions
of the issues involved. Recently, progress has also been achieved using WKB
techniques. These developments are reviewed in~\cite{Cc}.

An alternative to perturbation theory as a method to test proposed
boundary reflection amplitudes is to study their implications for the spectrum of
a theory on a finite interval, question (e) above. 
Integral equations (`of TBA [Thermodynamic Bethe Ansatz] 
type') encoding the ground-state
energy were first given in \cite{LMSS}. It turned out that these equations are not
always correct, and in \cite{DPTWa} modifications for situations where
they break down were found, along with
their generalisations to cover excited states. An important part of this work was
the testing of results against a generalisation of the so-called truncated
conformal space approach (TCSA) to models with boundaries. Further discussion of
these methods, and more
references, can be found in \cite{onept,parisg,parisr}, and in
G\'erard Watts' contribution to these proceedings. However the number of massive
models for which this programme has been completed remains small, and further
examples would be welcome. Progress on these issues is also
being made from the point of view of lattice models \cite{pearce}.

If the r\^oles of the space and time directions are swapped over, then the boundary
condition is replaced by an initial boundary {\em state}, $|B\,\rangle$ \cite{GZa}.
Space now runs from $-\infty$ to $+\infty$, and can be compactified to a circle
by imposing periodic boundary conditions. The inner product of $|B\,\rangle$
with the ground state for the theory on the circle is of particular interest,
being related to the so-called $g$-function of Affleck and Ludwig \cite{AL}. 
An alternative
set of integral equations, also of TBA type, can be used to study
the evolution of this quantity with system
size 
%\cite{blz,LMSS,gfl,AN}
[32,25,33,34]. These equations are, however, only fully
satisfactory when the bulk theory is massless \cite{gfl}. For the time
being it is an open
question whether modifications can be  found to describe the
flow of the $g$-function in the more general massive situation.

This has been a very rapid tour, giving only a flavour of the literature on boundary
integrable quantum field theories. To finish, I shall just mention three further
review articles. Reference \cite{Cd} is recommended for a more detailed treatment of the 
formal aspects of the subject,
while the articles \cite{Sb,Sc} are excellent starting-points for those
interested in the applications of
boundary integrability to concrete problems in condensed matter physics. 

\section*{\large\bf Acknowledgements}
%%%%%%%%%%%%%%%%%%%%%%%%%%%%%%%%%%%
%
I would like to thank 
Zal\'an Horv\'ath and Laci Palla for the invitation to discuss this subject at
the E\"otvos school, Ed Corrigan, Roberto Tateo and G\'erard Watts for helpful 
comments, and
the UK EPSRC for an Advanced Fellowship.
I was supported 
in part by a TMR grant of the
European Commission, reference ERBFMRXCT960012.
%
%
%%%%%%%%%%%%%%%%%%%%%%%%%%%%%%%%%%%w
%

\end{document}